\documentclass[11pt]{article}
\usepackage{fullpage}
\usepackage{epsfig}

\newtheorem{theorem}{Theorem}

\newtheorem{lemma}[theorem]{Lemma}

\newenvironment{proof}{\noindent{\bf Proof:}}{
\hspace*{\fill} \rule{2mm}{2mm} \vskip \belowdisplayskip}

\newenvironment{proofof}{\noindent{\bf Proof of}}{
\hspace*{\fill} \rule{2mm}{2mm} \vskip \belowdisplayskip}

\newcommand{\ve}{\varepsilon}

\newcommand{\cm}[1]{}

\newcommand{\ab}[1]{
{\left\vert {#1} \right\vert}
}

\newcounter{algline}

\newcommand{\nl}{\\ \thealgline \stepcounter{algline}}

\newenvironment{algtab2}{\begin{minipage}[t]{.04\linewidth}~
\end{minipage} \begin{minipage}[t]{0.92\linewidth}\vspace{-.06in}}{\vspace{-.0in
}\end{minipage}}

\newenvironment{boxfig}[1]{\begin{figure*}[htb]\fbox{\begin{minipage}{\linewidth}
                        \vspace{1em}
                        \makebox[0.025\linewidth]{}
                        \begin{minipage}{0.95\linewidth}
                        #1
                        \end{minipage}
                        \end{minipage}}}{\end{figure*}}

\begin{document}
\title{Source Routing and Scheduling in Packet Networks~\thanks{Partially 
supported by DIMACS funding. A preliminary version of this paper
appeared in the Proceedings of the 42th IEEE Annual Symposium on
Foundations of Computer Science, FOCS 2001.}}
\author{Matthew Andrews~\thanks{Bell Laboratories. andrews@research.bell-labs.com.}
\and
Antonio Fern\'andez~\thanks{GSyC, ESCET, Universidad Rey Juan Carlos, Spain. 
anto@gsyc.escet.urjc.es.}
\and
Ashish Goel~\thanks{Department of Computer Science, University of
Southern California.  agoel@cs.usc.edu.}
\and
Lisa Zhang~\thanks{Bell Laboratories. ylz@research.bell-labs.com.}}
\maketitle
\thispagestyle{empty}

\begin{abstract}

We study {\em routing} and {\em scheduling} in packet-switched
networks.  We assume an adversary that controls the injection time,
source, and destination for each packet injected.  A set of paths for
these packets is {\em admissible} if no link in the network is
overloaded.  We present the first on-line routing algorithm that finds
a set of admissible paths whenever this is feasible.  Our algorithm
calculates a path for each packet as soon as it is injected at its
source using a simple shortest path computation. The length of a link
reflects its current congestion.  We also show how our algorithm can
be implemented under today's Internet routing paradigms.

When the paths are known (either given by the adversary or computed as
above) our goal is to schedule the packets along the given paths so
that the packets experience small end-to-end delays.  The best
previous delay bounds for deterministic and distributed scheduling
protocols were exponential in the path length.  In this paper we
present the first deterministic and distributed scheduling protocol
that guarantees a polynomial end-to-end delay for every packet.

Finally, we discuss the effects of combining routing with scheduling.
We first show that some unstable scheduling protocols remain unstable
no matter how the paths are chosen.  However, the freedom to choose paths
can make a difference.  For example, we show that a ring with parallel
links is stable for all greedy scheduling protocols if paths are
chosen intelligently, whereas this is not the case if the adversary
specifies the paths.
\end{abstract}

\section{Introduction}
\label{s:intro}

Two of the most important problems in the control of packet-switched
networks are {\em routing} and {\em scheduling}.  The goal of routing
is to assign a path to a packet from its source to its destination.
The goal of scheduling is to deal with the {\em contention} that
occurs when two or more packets wish to cross a link simultaneously.
Each link must have a {\em scheduler} that resolves this contention
by deciding which packet to advance.

The scheduling problem typically assumes that the paths of the packets
are given as part of the input.  The goal is then to schedule the
packets along their paths in such a way that they all reach their
destinations in a short time.  Much recent work has focused on the
{\em Adversarial Queueing Model},
e.g.~\cite{BorodinKRSW96,AndrewsAFKLL96,Gamarnik98}.  We follow
their convention and assume that all packets are unit size and each link
processes one packet per time step.  In this Adversarial Queueing
Model, the adversary chooses the injection time, source, destination,
and route for each packet injected.  A sequence of injections is
called $(w,r)$-{\em admissible} for a window size $w$ and injection
rate $r<1$, if in any time interval of $T\ge w$ the total number of
packets injected into the network whose paths pass through any link
$e$ is at most $Tr$.  These paths are also called $(w,r)$-{\em
admissible}.  Previous work has examined the performance of a number
of simple scheduling protocols in this model.  A packet scheduling
protocol is said to be {\em universally stable} if it guarantees
bounded buffer sizes and packet transmission delays for any
$(w,r)$-admissible injections. In~\cite{AndrewsAFKLL96} it was proved
that several natural protocols (Longest-In-System, Shortest-In-System,
Furthest-To-Go) are universally stable, whereas several others
(First-In-First-Out, Last-In-First-Out, Nearest-To-Go) are not.

In this paper we study both routing and scheduling.  The adversary no
longer specifies the route of each packet; it merely specifies the
source and destination.  However, we are guaranteed 
that $(w,r)$-admissible paths for the injections do exist.  The
problem is now two-fold. We first need to find some $(W,R)$-admissible
paths, possibly for a different window size $W$ and a different $R<
1$. These admissible paths combined with a universally stable
scheduling scheme, such as the ones in~\cite{AndrewsAFKLL96} or the
one presented in Section~\ref{s:polydelay} of this paper, result in a
universally stable protocol for routing and scheduling.

\subsection{Source Routing for Stability}

\paragraph{Our result.}
In Section~\ref{s:basic-routing} of the paper we present the first
online algorithm for assigning admissible routes to packets.  If the
adversary can assign $(w,r)$-admissible routes, then our algorithm
finds a set of $(W,R)$-admissible routes where $R\in (r,1)$ is of our
choice and $W\ge w$ is determined by the choice of $R$. Hence, if the
parameter of merit is the window size $w$, then our algorithm is a
$W/w$-approximation algorithm (modulo a small increase in the rate).
Moreover, our algorithm is online in that it assigns routes to packets
as soon as they are injected into the network.  Hence it can also be
regarded as a $W/w$-competitive algorithm for this problem.  This is
the first approximation algorithm/competitive algorithm for this
problem.  Once the routes are chosen, we can use any ``good''
scheduling protocol in the Adversarial Queueing Model.

Our algorithm is based on the $\ve$-approximation algorithm for
fractional maximum multicommodity concurrent flow given by Garg and
K\"onemann~\cite{GargK98}, which in turn builds upon the work of
Plotkin, Shmoys, and Tardos~\cite{pst:pack95} and
Young~\cite{y:round95}. In the maximum multicommodity concurrent flow
problem, the demands for each commodity remain constant as the
algorithm progresses. In our setting, the demands between
source-destination pairs correspond to the packets injected by the
adversary, which can change over time. Even though the algorithm of
Garg and K\"onemann~\cite{GargK98} is an {\em offline} algorithm that
assigns {\em fractional} paths to a fixed set of commodities, in our
setting we are able to convert it into an {\em online} algorithm that
assigns an {\em integral} path to each packet as soon as it is
injected.

\paragraph{Implementation under Internet routing paradigms.}


At a high level, our algorithm works as follows.  Each link maintains
a measure of {\em congestion} that represents how many packets have
been routed through it in the recent past.  Packets are then routed on
shortest paths with respect to this congestion measure.  Hence we need
a mechanism for distributing congestion information from the links to
the source nodes.  We also need a mechanism by which a source node can
inform a link whenever it routes a packet through that link. 

The first requirement could be satisfied by something akin to the OSPF
(Open Shortest Path First) link state flooding protocol. (See
e.g.~\cite{Keshav97}.) This is a protocol that is used for flooding
link state information to the nodes in a network so that packets may
be routed along shortest paths.  The second requirement may be
satisfied by the MPLS (Multi-Protocol Label Switching) protocol that
is gaining increasing acceptance in the Internet. (See e.g.\
\cite{RFC3031}.)  With this protocol a source node can compute an
explicit route to each destination and then distribute a label for the
route to each of the links that comprise the route.  In combination
with this label distribution the source can also specify how much
traffic it is going to send on the route.

In Section~\ref{s:basic-routing} we first assume that this control
information is transmitted instantaneously and does not contribute to
the congestion in the network.  We then consider a model in which the
control information is transmitted in-band through the network and
must contend with the data traffic.

\paragraph{Relation to previous work.}
Routing and scheduling as a combined problem has been studied
in the past.  For example, Aiello et al.\ presented a distributed
algorithm~\cite{AielloKOR98} motivated by the Awerbuch-Leighton
multicommodity flow algorithm~\cite{AwerbuchL94}.
In~\cite{Gamarnik99} Gamarnik gave a solution based on an
approximation algorithm for static routing.
However, both these algorithms require a dependence between how a
packet is routed and how it is scheduled. Hence, their routing schemes
only work in association with their specific scheduling schemes, but
not with generic scheduling algorithms.  Neither routing algorithm can
be used to provide packets with admissible paths at injection time.
Using networking terminology, these routing algorithms correspond to
{\em active routing}~\cite{Braden}, where intermediate routers need to
actively participate in determining routes for each individual packet.
In contrast, our algorithm corresponds to {\em source
routing},
where the entire path of a packet is known at the source.


\subsection{Deterministic Distributed Scheduling with Polynomial Delays}

In Section~\ref{s:polydelay} of the paper we study the scheduling
problem in isolation assuming that $(w,r)$-admissible paths are given.
In recent years, a number of scheduling algorithms have been proposed
that guarantee {\em network stability}, i.e.\ the number of packets in
the network remains bounded and the end-to-end delay experienced by
packets remains bounded.  For example, the Longest-In-System protocol
that always gives priority to the packet injected into the system
earliest, was shown in \cite{AndrewsAFKLL96} to guarantee a delay
bound of $O(w/(1-r)^{d_{\max}})$, where $d_{\max}$ is the maximum
length of a path assigned to any packet.  Note however, that this
bound is exponential in $d_{\max}$.  It has been an open problem
whether or not any deterministic, distributed scheduling protocol has
a polynomial delay bound in the Adversarial Queueing Model.  Indeed,
\cite{AndrewsAFKLL96} remarked that ``it is of considerable interest
to determine whether such a protocol exists''.

A {\em randomized} protocol based on Longest-In-System can guarantee
that each packet experiences a delay of $poly(w,1/(1-r),d_{\max},\log
m)$ with high probability~\cite{AndrewsAFKLL96}, where $m$ is the
number of links in the network.  In essence, for most of the time the
protocol is successful and keeps all delays small.  However, even if
the failure probability is small, if the algorithm is run for an
extended period of time then the algorithm is likely to make some
random choices that are bad.  This causes packets to violate the delay
bound. Moreover, if one packet violates the delay bound then other
packets injected along the same path at similar times are also likely
to violate the delay bound.  Hence, all of the packets that make up a
single file transfer could be excessively delayed.  Although this
randomized protocol can be derandomized in a centralized manner it
seems hard to convert it into a deterministic, {\em distributed}
protocol.  This is because the ``success condition'' involves packets
injected at multiple source nodes and hence it cannot be verified
locally.

\paragraph{Our result.} 
In Section~\ref{s:polydelay} we present the first deterministic,
distributed scheduling protocol with a polynomial delay bound.  It
guarantees that {\em all} packets reach their destination within
$poly(w,1/(1-r),m)$ steps of their injection.  We start by presenting
a randomized protocol in which the ``success condition'' can be
verified at the source nodes independently.  This allows us to
derandomize the protocol in a distributed fashion.

\subsection{The Effects of Combining Source Routing with Scheduling}

In the final part of the paper we consider the following question: Is
it possible for unstable scheduling protocols to become stable if
paths can be chosen by a routing algorithm as opposed to being
dictated by the adversary? We first present a network and a sequence
of packet injections such that regardless of how the routes for these
packets are chosen, many greedy protocols (including FIFO) remain
unstable. Thus, we cannot hope to achieve stability using FIFO even if
we have the freedom to choose routes. However, we also present an
example in which the ability to select the routes does make a
difference. We show that in a ``ring'' with multiple parallel links,
if we are allowed to choose the routes intelligently then we can
ensure that all greedy scheduling protocols are stable.  However, if
the adversary dictates the routes then many scheduling protocols
(including FIFO) are unstable.

\subsection{Other Related Work}

Much traditional work on routing focuses on the problem of routing
{\em flows} online, e.g.~\cite{AwerbuchAP93,AwerbuchAPW94}.  Each flow
requests a bandwidth from a source to a destination and we must choose
a path for each accepted flow without violating any link capacity.
The goal is to maximize the on-line acceptance rate.  However, this
work does not consider packet-level behavior.

The problem of choosing routes for a fixed set of packets was studied
by Srinivasan and Teo~\cite{SrinivasanT97} and Bertsimas and
Gamarnik~\cite{BertsimasG99}.  For example, \cite{SrinivasanT97}
presents an algorithm that minimizes the congestion and dilation of
the routes up to a constant factor.  This result complemented the
paper of Leighton, Maggs and Rao~\cite{LeightonMR88} which showed that
packets could be scheduled along a set of paths in time
$O($congestion$+$dilation$)$.


\section{Source Routing for Stability}
\label{s:basic-routing}


For convenience we use the following weaker notion of admissibility in
this section. We say that a set of packet paths is {\em weakly}
$(w,r)$-{\em admissible} if we can partition time into windows of
length $w$ such that for each window {\em in the partition} and each
link $e$, the number of paths that pass through $e$ and correspond to
packets injected during the window is at most $wr$.  However, this
distinction is not important due to Lemma~\ref{l:admissible}.
Moreover, all of the delay bounds that have been derived in the past
for the Adversarial Queueing Model apply to weakly $(w,r)$-admissible
paths.
\begin{lemma}
\label{l:admissible}
If a set of paths is $(w,r)$-admissible then it is also weakly
$(w,r)$-admissible.  Conversely, weak $(w,r)$-admissibility implies
$(w',r')$-admissibility for some $w'\ge w$ and $r'\in [r,1)$.
\end{lemma}
\begin{proof}
Suppose the injections are weakly $(w,r)$-admissible.  We show that they are
 $(w',r')$-admissible for $r'=(1+r)/2$ and $w'=4wr/(1-r)$.
For any $T\ge w'$, let $T$ be in the range of $[nw, (n+1)w)$ where $n$
is an integer at least $4r/(1-r)$. Due to weak admissibility and our choices
of $n$, $T$ and $r'$, the number of injections during $T$ steps for
any link $e$ is at most,
$$
(n+2)rw \le nw (1+r)/2 \le Tr'.
$$
The other direction is trivial.
\end{proof}

We assume an adversary that injects weakly $(w,r)$-admissible packets
into the network\footnote{In fact, as will be seen later, we only need
to assume that the adversary can choose {\em fractional} paths that
are weakly $(w,r)$-admissible.}.  Our aim is to choose weakly
$(W,R)$-admissible routes for these packets where $R\in (r,1)$ is of
our choice and $W\ge w$ is determined by the choice of $R$.

\subsection{The Basic Routing Protocol}
We first assume that control information is communicated
instantaneously. Whenever a source node chooses a route for a packet,
this information is instantaneously transmitted to all the links on
the route.  Whenever the congestion on a link changes, this fact is
instantaneously transmitted to all the source nodes. Later on we relax
these assumptions. As mentioned in the Introduction, the algorithm is
based on the Garg-K\"onemann offline approximation algorithm for
fractional maximum concurrent flow.  However, in our setting we can
convert it into an {\em online} algorithm that chooses {\em integral}
paths for the packets.  

\begin{boxfig}{
\begin{algtab2}
\setcounter{algline}{1}
\begin{tabbing}
123\=6789\=9012\=12345\=67890\=12345\=67890\=12345\= \kill
Find routes.\nl
\> Initialize $c(e) = \delta$, $\forall e$\nl
\> for the $i$th window, $i=1,\dots,t$ \nl
\>\> for each packet injected during $i$th window \nl
\>\>\> $p \leftarrow$ least congested route under $c$ (i.e.\ shortest
path with respect to $c$)\nl
\>\>\> $c(e)\leftarrow c(e)(1+\mu/w)$, $\forall e\in p$ 
\end{tabbing}
\end{algtab2}
}
\caption{Procedure to find routes for packets injected during one phase.}
\label{f:ell}
\end{boxfig}

\paragraph{Protocol.}
We route every packet injected along the path whose total congestion
is the smallest under the current congestion function $c(\cdot)$,
i.e.\ we route along shortest paths with respect to $c(\cdot)$.
Initially, the congestion along every link is set to $\delta$ where
$\delta$ is defined in~(\ref{eq:delta}).  
For every link $e$ along the chosen route,
its congestion $c(e)$ is updated to $c(e) (1+\mu/w)$ where $\mu$
is defined in~(\ref{eq:mu1}).  We reset the congestion of every link to
its initial value of $\delta$ at the beginning of each {\em phase}.  A
phase terminates in $t$ windows of $w$ steps, where $t$ is an integer
defined in~(\ref{eq:t}).  Figure~\ref{f:ell} illustrates the procedure
for one phase.
The values of $\mu$, $\delta$ and $t$ are defined as follows.  Let $m$
be the number of links in the network. For any $R\in (r,1)$ of our
choice, let
\begin{eqnarray}
\mu &=& 1-\left({r\over R}\right)^{1/3} \label{eq:mu1} \\
\delta &=& \left({1-r\mu}\over m\right)^{1/r\mu} \label{eq:delta}\\
t & = & \left\lfloor{{1-r\mu}\over {r\mu}}\ln {{1-r\mu}\over{m\delta}}
\right\rfloor + 1 \label{eq:t}
\end{eqnarray}
Our objective is to show,
\begin{theorem}
\label{t:congestion}
For all packets injected during one phase, at most $tw R$ of their
routes chosen by our procedure go through the same link.  In other
words these routes are weakly $(tw,R)$-admissible.
\end{theorem}
\paragraph{Analysis.}
To prove Theorem~\ref{t:congestion} let us examine an integer program
formulation for routing the set of packets injected during a
window of $w$ time steps.  Let $P_j$ be the set of possible routes for
the $j$th packet, and let variable $x_j(p)\in\{0,1\}$ indicate whether
or not route $p \in P_j$ is chosen for packet $j$.  The following
linear relaxation of the integer program (LP) has an optimal solution
$\lambda \ge 1$ since the injections are $(w,r)$-admissible. We
present both the primal and the dual.
$$
\begin{array}{rcl}
     & \mbox{Primal} & \\
     & \max \lambda & \\ 
s.t. & & \\
     & \sum_{p\in P_j} x_j(p) \ge \lambda  & \forall j \\
     & \sum_j \sum_{p: e\in p, p\in P_j} x_j(p) \le rw & \forall e \\
     & x_j(p) \ge 0 & \forall j, \forall p\in P_j 
\end{array}
$$
$$
\begin{array}{rcl}
     & \mbox{Dual} & \\
     & \min \sum_e rw\cdot c(e) &   \\
s.t. & & \\
     & \sum_{e\in p} c(e) \ge z(j)  &  \forall j, \forall p\in P_j \\
     & \sum_j z(j) \ge 1 & \\
     & c(e) \ge 0 & \forall e\\
     & z(j) \ge 0 & \forall j
\end{array}
$$
For any non-negative congestion function $c(\cdot)$, let $D = \sum_e
c(e)$ be the total congestion of all links.  For packet
$j$ let $q_j$ be the least congested
path in terms of $c$.  We use $\alpha=\sum_j \sum_{e\in q_j}
c(e)$ to represent the total congestion of these least congested
paths. It can be shown that the dual is equivalent to,
$$
\min_c \mbox{  }rw \cdot D/ \alpha.
$$
The congestion found at the end of window $i$ by our protocol (see
Figure~\ref{f:ell}) defines a valid solution to this reformulated dual
for window $i$.  We exploit this connection to prove
Theorem~\ref{t:congestion}.  The key here is to bound the total link
congestion since the link congestion increases only when a path goes
through it.  In particular, the following three lemmas show that the
total link congestion is no more than 1 at the end of a phase.  Let
$c_i(e)$, $D_i$ and $\alpha_i$ represent the values of $c(e)$, $D$ and
$\alpha$ at the end of the $i$th window.

\begin{lemma}
\label{l:duality}
$D_i/\alpha_i\ge 1/rw$ for $1\le i\le t$.
\end{lemma}
\begin{proof}
Since the injections are $(w,r)$-admissible, the primal LP for window
$i$ has $\max\lambda \ge 1$.  Since the congestion $c_i$ found by our
protocol defines a dual solution, our lemma follows from duality.
\end{proof}
\begin{lemma}
\label{l:di}
$D_i \le {D_{i-1}\over {1-r\mu}}$.
\end{lemma}
\begin{proof}
It suffices to show $D_i\le D_{i-1} + \alpha_i \cdot\mu /w$ since
$D_i/\alpha_i\ge 1/rw$ by Lemma~\ref{l:duality}.  Let $c_{ij}$ be the
congestion function after routing the $j$th packet injected during the
$i$th window and let $D_{ij}$ be defined in terms of $c_{ij}$.
Suppose path $p_j$ is chosen for the $j$th packet injected during the
$i$th window.  
By definition we have,
\begin{eqnarray*}
D_{ij} &=& \sum_e c_{ij}(e) \\
       &=& \sum_{e\notin p_j} c_{i,j-1}(e) + 
             \sum_{e\in p_j} c_{i,j-1}(e) (1+\mu/w) \\
       &=& D_{i,j-1} + \sum_{e\in p_j} c_{i,j-1}(e) \cdot\mu /w.
\end{eqnarray*}
Now we repeatedly apply the recurrence above.  We also observe that
the congestion function $c$ only increases.  Hence, if $q_j$ is the
least congested path for $j$ under $c_i$ then $\sum_{e\in p_j}
c_{i,j-1}(e)$ is necessarily no more than $\sum_{e\in q_j}
c_{i}(e)$.  
(We emphasize that $p_j$ and $q_j$ may be two different
paths.  The path $p_j$ is least congested with respect to
$c_{i,j-1}$ and $q_j$ is least congested with respect to $c_i$.) 
We have,
\begin{eqnarray*}
D_i &=& D_{i-1} + \sum_j \sum_{e\in p_j} c_{i,j-1}(e) \mu/w \\
&\le& D_{i-1} + \alpha_i\cdot \mu/w.
\end{eqnarray*}
\end{proof}

\begin{lemma}
$D_t \le 1$.
\end{lemma}
\begin{proof}
By definition $D_0 = m\delta$ where $m$ is the number of links in
the network.  By applying Lemma~\ref{l:di}, we have,
\begin{eqnarray*}
D_t &\le& {{m\delta}\over {(1-r\mu)^t}} \\
    &=& {m\delta\over {1-r\mu}}\left(1+{r\mu\over{1-r\mu}}\right)^{t-1} \\
    &\le&  {m\delta\over {1-r\mu}} e^{r\mu(t-1)\over {1-r\mu}} \\
    &\le& 1.
\end{eqnarray*}
The second inequality follows from $1+x \le e^x$ for $x\ge 0$.  The
last inequality follows from the definition of $t$ in~(\ref{eq:t}).
\end{proof}

We are now ready to prove Theorem~\ref{t:congestion}.

\begin{proofof}{ \bf Theorem~\ref{t:congestion}:}
Consider any link $e$.  For every $w$ paths routed though $e$, the
congestion of $e$ is increased by a factor at least $1+\mu$.  Initially,
$c_0(e) = \delta$.  Since $D_t\le 1$, $c_t(e) \le 1$. Hence,
the total number of paths that are routed through $e$ in a phase
is at most $w\log_{1+\mu}{1/\delta}$.
It suffices to show that this quantity is no more than $wt R$.
\begin{eqnarray*}
{{w\log_{1+\mu}{1/\delta}} \over {wt R}} & \le &
{\ln {1/\delta} \over \ln (1+\mu)} \cdot {r\mu \over{1-r\mu}} 
\cdot {1\over \ln {{1-r\mu}\over{m\delta}}} \cdot {1\over R} \\
& = & {r\over R}\cdot {\mu \over {\ln(1+\mu)(1-r\mu)^2}} \\
& \le & {r\over R}\cdot (1-\mu)^{-3} \\
& = & 1.
\end{eqnarray*}

The first inequality and the first equality follow from the
definitions of $t$ and $\delta$ respectively.  The second inequality
follows from the fact that $r<1$ and $\ln (1+\mu) \ge \mu -\mu^2/2$.
The last equality follows from the definition of $\mu$. Our proof is
complete.
\end{proofof}

\begin{boxfig}{
\begin{algtab2}
\setcounter{algline}{1}
\begin{tabbing}
123\=6789\=9012\=12345\=67890\=12345\=67890\=12345\= \kill
Find routes.\nl
\> Initialize $c(e) = \delta$, $\forall e$\nl
\> for $i$th window, $i=1,\dots,t$ \nl
\>\> for each packet injected during $i$th window \nl
\>\>\> $p \leftarrow$ least congested route under $c$ \nl
\>\> $c(e)\leftarrow c(e)(1+N_i(e)\cdot \mu/w)$. 
\end{tabbing}
\end{algtab2}
}
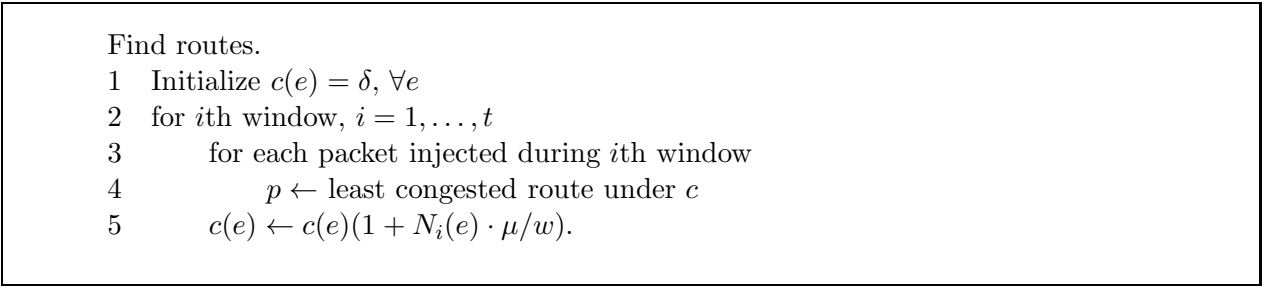
\caption{Procedure to find routes for packets injected during one phase with
fewer updates.}
\label{f:newell}
\end{boxfig}

\subsection{Routing with Less Frequent Updates}
\label{s:batch}
In this section we show that Theorem~\ref{t:congestion} still holds
even if the congestion function $c$ is updated less frequently.  In
particular, we only update the congestion at the end of each window,
not for each packet injection.  Hence the source nodes only need to
communicate with the links at the end of each window.  For this new
protocol we redefine $\mu$ to be
\begin{eqnarray}
\label{eq:mu2}
\frac{1}{m}\left(1-\left({r\over R}\right)^{1/3}\right). 
\end{eqnarray}
Suppose $N_i(e)$ packets are routed through link $e$ during the $i$th
window, then we update $c(e)$ to $c(e) (1+N_i(e) \cdot\mu /w)$.  See
Figure~\ref{f:newell}.

We prove that Theorem~\ref{t:congestion} remains true.  We first show
that Lemma~\ref{l:di} still holds.  As before, we show $D_i\le D_{i-1}
+ \alpha_i \cdot\mu /w$.  For any packet $j$ injected during the $i$th
window, let $p_j$ be the path chosen for $j$.
\begin{eqnarray*}
D_{i} & = & \sum_e c_{i}(e) \\
      & = & \sum_e c_{i-1}(e) (1+N_i(e)\cdot\mu/w) \\
      & = & D_{i-1} + \sum_e c_{i-1}(e) N_i(e)\cdot\mu/w \\
      & = & D_{i-1} + \sum_j\sum_{e\in p_j} c_{i-1}(e) \cdot \mu/w \\
      & \le & D_{i-1} + \alpha_i \cdot \mu /w
\end{eqnarray*}
Hence $D_t\le 1$. Now, for every $mw$ paths routed through $e$, the
congestion on $e$ is increased by a factor at least $1+m\mu$.  Therefore
the congestion on any link at the end of a phase is at most,
\begin{eqnarray*}
{{mw\log_{1+m\mu}{1/\delta}} \over {wt R}} & \le &
{\ln {1/\delta} \over \ln (1+m\mu)} \cdot {r\mu \over{1-r\mu}} 
\cdot {1\over \ln {{1-r\mu}\over{m\delta}}} \cdot {1\over R} \\
& = & {r\over R}\cdot {m\mu \over {\ln(1+m\mu)(1-r\mu)^2}} \\
& \le & {r\over R}\cdot (1-m\mu)^{-3} \\
& = & 1,
\end{eqnarray*}
with the revised definition of $\mu$ in~(\ref{eq:mu2}).

\subsection{Implementation Using In-band Signaling}

In the previous sections we assumed that sources can communicate with the
links on their chosen routes via instantaneous setup messages.  In
turn, we also assumed that the links can instantaneously broadcast
their congestion to the sources.  In this section, we first
extend our result in Section~\ref{s:batch} to the case where each of these
communications takes $\tau$ time steps.  We then give an upper bound
on $\tau$ for which the communication may be carried out in-band using
packets transmitted through the network.

Assume without loss of generality that $w>2\tau$ (since admissibility
for a small window implies admissibility for a large window).  Each
source only updates the link congestion at the end of every window.
Since the congestion does not change during a window, all the packets
for a given source-destination pair $(s,t)$ are routed along the {\em
same path} $p$.  At the end of window $[w(i-1),wi)$ a {\em control
packet} is sent along path $p$ that contains the number of
$(s,t)$-packets injected during window $[w(i-1),wi)$.  This packet
takes time $\tau$ to traverse the path.  Hence, at time $wi+\tau$,
each link can update its congestion due to all the packets injected
during $[w(i-1),wi)$.  Then by time $wi+2\tau\le w(i+1)$ this new
congestion can be distributed via control packets to all the sources.

Note that at the end of window $[wi,w(i+1))$, every link has updated
its congestion according to the injections in window
$[w(i-1),wi)$. The exact form of this update is as follows.  Let
$N_i(e)$ be the number of packets routed through $e$ that were
injected during $[w(i-1),wi)$.  Let $c_i(e)$ be the congestion of $e$
at the end of window $[w(i-1),wi)$.  We update $c_i(e)$ by,
$$
c_{i+1}(e) = c_{i}(e) + c_{i-1}(e)N_i(e)\cdot\mu/w,
$$
for 
\begin{eqnarray}
\label{eq:mu3}
\mu = \frac{1}{2m}\left(1-\left({r\over R}\right)^{1/3}\right).
\end{eqnarray}
To show that Theorem~\ref{t:congestion} remains true, we observe,
\begin{eqnarray*}
D_{i+1} & = & \sum_e c_{i+1}(e) \\
      & = & \sum_e c_{i}(e) + c_{i-1}N_i(e)\cdot\mu/w \\
      & = & D_{i} + \sum_e c_{i-1}(e) N_i(e)\cdot\mu/w \\
      & = & D_{i} + \sum_j\sum_{e\in p_j} c_{i-1}(e) \cdot \mu/w \\
      & \le & D_{i} + \alpha_{i,i+1} \cdot \mu /w.
\end{eqnarray*} 
Here $\alpha_{i,i+1}$ is the sum of the congestion along the paths
chosen for packets injected during $[w(i-1),wi)$ with respect to
$c_{i+1}(e)$.  This is sufficient to imply $D_t \le 1$. Note also that
for every $2mw$ (non-control) packets routed through a link, the
congestion function of the link increases by at least a factor
$1+2m\mu$. The remainder of the analysis follows through for
the revised definition of $\mu$ in~(\ref{eq:mu3}).
 
To ensure that the transmission time of the control packets is upper
bounded, the scheduling protocol always gives priority to control
packets.  Observe that a total of at most $n^2 + mn$ control packets
can be sent out during one window, where $m$ is the number of links
and $n$ is the number of nodes in the network.  If we let $\tau = n^3
+ mn^2$, the transmission of a control packet takes at most $\tau$
time steps.  Without loss of generality we assume that $w\ge 2\tau$
and $w(1-r)/2 \ge n^2 + mn$.  The latter condition ensures that
together with the control packets the injections are
$(w,(1+r)/2)$-admissible.

\section{A Scheduling Protocol with Polynomial Delay Bounds}
\label{s:polydelay}

In this section we assume that $(w,r)$-admissible paths are known
(either given by the adversary or computed as in
Section~\ref{s:basic-routing}).  Hence, in order to achieve network
stability we can use any of the scheduling protocols that are known to
be stable for Adversarial Queueing.  However, the best previous delay
bounds known for distributed, deterministic protocols are exponential
in the maximum packet path length.  In this section we present a
deterministic, distributed scheduling protocol with a polynomial delay
bound.

In \cite{AndrewsAFKLL96} a randomized protocol was presented for which
the delay bound is 
$O(\frac{d_{\max}}{\ve}\log m)$
with high probability,
where $\ve = 1-r$ and $d_{\max}$ is the length of the longest simple
path in the network.  This protocol is hard to derandomize because its
success depends on a condition that can only be checked globally.  In
this section we first present a new randomized protocol and then show
how to derandomize it in a distributed manner. The key idea of this
protocol is that the conditions that determine the ``success'' of the
protocol only depend on packets that share the same initial link.
This allows derandomization in a distributed manner.

Our new randomized protocol is defined in terms of two parameters $M$
and $T$ which are defined below.  We partition time into intervals of
length $M$, which we call $M$-{\em intervals}.  We save up all packets
that are injected into the network during each $M$-interval and then
schedule these packets during the next $M$-interval.  We give each
packet a deadline for every link on its path. Our goal is to make sure
that no more than $T$ packets have a deadline for link $e$ during any
time interval of length $T$.  If this condition holds then we are able
to bound the end-to-end delay experienced by a packet.

\paragraph{Randomized protocol.} 
For a packet $p$ injected during an $M$-interval $[(\gamma-1)M,\gamma
M)$ for an integral $\gamma$, let us suppose its path is
$e_0,e_1,\ldots,e_{d_p}$.  We define a deadline $\tau_k^p$ for $p$ at
link $e_k$ as follows.  We choose the initial deadline $\tau^p_0$
uniformly at random from $[\gamma M+T,(\gamma+1)M-d_{\max}T)$. We then
define the remaining deadlines inductively by
$\tau_{k+1}^p=\tau_k^p+T$.  Our protocol always gives priority to the
packet with the smallest deadline at each link.
We define $M$ and $T$ such that,
\begin{eqnarray}
T &=& \frac{36m}{\ve^3}\log (2Mm^2) \label{eq:T},\\
M &\ge& \max\left\{\frac{1-\ve/2}{\ve/6}(d_{\max}+1)T, w\right\}
\label{eq:M}.
\end{eqnarray}
These properties are satisfied for,
$$
M=O\left(\frac{d_{\max}m}{\ve^4} \log {\frac{m}{\ve}} + w\right).
$$
When a packet meets its deadlines, it reaches its destination
within $2M$ steps. 

\paragraph{Analysis.}
Our objective is to show that all packets injected during a given
$M$-interval meet all their deadlines with a constant probability.
Lemma~\ref{l:undercongested} gives a sufficient condition for all
deadlines to be met.  For any packet $p$ and link $e$ let
$X^{p,e}_{[t,t+T)}=1$ if $e$ is the $k$th link on packet $p$'s path
and $\tau^p_k$ lies in the time interval $[t,t+T)$.  Let
$X^{p,e}_{[t,t+T)}=0$ otherwise.
\begin{lemma}
\label{l:undercongested}
If
$
\sum_p X^{p,e}_{[t,t+T)} \le T
$
for all $t$ and all links $e$, then all packets meet all their
deadlines.
\end{lemma}
\begin{proof}
Suppose not. Let $p$ be a packet that misses its $k$th deadline
$\tau_k^p$ and suppose that no deadline earlier than $\tau_k^p$ is
missed.  Then $p$ has arrived at its $k$th link $e_k$ by time
$\tau_k^p-T$. (This is true regardless of whether $e_k$ is the initial
link of $p$ or not.)  By our assumption that $\tau_k^p$ is the first
deadline that is missed, all the packets with deadlines for $e_k$ that
are earlier than $\tau_k^p-T+1$ meet those deadlines.  Therefore, the
only packets that block packet $p$ in the interval
$[\tau^p_k-T+1,\tau^p_k]$ have deadlines in the interval
$[\tau^p_k-T+1,\tau^p_k]$.  By the assumption in the statement of
the lemma there are at most $T-1$ such packets (excluding $p$).
Therefore packet $p$ is served by link $e_k$ at time $\tau^p_k$ or
earlier. This is a contradiction.
\end{proof}

Given Lemma~\ref{l:undercongested} we show,
\begin{lemma}
\label{l:whp}
Consider packets injected during an $M$-interval, $[(\gamma-1)M,\gamma
M)$.  The number of deadlines from these packets on any link $e$
during any interval $[t,t+T)$ is at most $T$ with a constant
probability.
\end{lemma}
\begin{proof}
We use a Chernoff bound to prove the number of deadlines is small.
Let $S_{e_0,e}^{\gamma}$ be the set of packets
injected into the network during the interval $[(\gamma-1)M,\gamma M)$
that have $e_0$ as their initial link and that have link $e$ on their
path.  The expected number of deadlines is,
$$
E\left[\sum_{p\in S_{e_0,e}^\gamma} X^{p,e}_{[t,t+T)}\right] \le 
{|S^\gamma_{e_0,e}|\over {M-(d_{\max}+1)T}}T.
$$
When $|S^\gamma_{e_0,e}|$ is large, the expectation is large and the
argument is straightforward. However, for small $|S^\gamma_{e_0,e}|$
a direct application of the Chernoff bound may not suffice.  To rectify
this, let us define a new quantity,
$$
\beta^\gamma_{e_0,e} = 
{M\over {M-(d_{\max}+1)T}}\max\{ |S^\gamma_{e_0,e}|/M, \ve/3m\}.
$$
The quantity $\beta$ has the following properties. 
\begin{enumerate}
\item
$\beta^\gamma_{e_0,e} \ge \ve/3m$;
\item 
$\sum_{e_0}\beta^\gamma_{e_0,e} \le 
\frac{M}{M-(d_{\max}+1)T}((1-\ve)+m\ve/3m) 
\le \frac{1-\ve/2}{1-2\ve/3}(1-2\ve/3) \le 1-\ve/2.
$
\end{enumerate}
The second property follows from the requirement of $M$
in~(\ref{eq:M}) and the admissibility of the paths.  
Our lemma follows if we show that the following holds
with constant probability,
{\small
\begin{eqnarray}
\label{eq:beta}
\sum_{p\in S_{e_0,e}^\gamma}X^{p,e}_{[t,t+T)} 
\le (1+\ve/2)\beta^\gamma_{e_0,e}T,
\forall e_0, e \mbox{ and } \forall [t,t+T).
\end{eqnarray}
}
If the above holds, the number of deadlines on link
$e$ in the interval $[t,t+T)$ is at most
$(1+\ve/2)\sum_{e_0}\beta^\gamma_{e_0,e}T$, which is less than $T$ due
to the second property of $\beta$.  We have,
\begin{eqnarray}
Pr\left[\sum_{p\in S_{e_0,e}^\gamma }X^{p,e}_{[t,t+T)} > 
(1+\ve/2)\beta^\gamma_{e_0,e}T\right] \nonumber
&\le & \frac{\prod_pE[(1+\ve/2)^{X^{p,e}_{[t,t+T)}}]} 
{(1+\ve/2)^{(1+\ve/2)\beta^\gamma_{e_0,e}T}} \label{eq:chernoff} \\
&\le& \exp(-\ve^2\beta^\gamma_{e_0,e}T/12) \nonumber \\
&\le& {1\over{2Mm^2}}.\label{eq:prob}
\end{eqnarray}
The first inequality is due to a Chernoff bound.  The second
inequality holds since $E[\sum_{p\in S_{e_0,e}^\gamma
}X^{p,e}_{[t,t+T)}]$ $\le \beta^\gamma_{e_0,e}T$ and $1+x \le e^x$ for
$x \ge 0$. The third inequality follows from the definition of $T$
in~(\ref{eq:T}) and the fact that $\beta^\gamma_{e_0,e}\ge \ve/3m$.
By taking a union bound over all links $e_0$, $e$ and all intervals
$[t,t+T) \subseteq [\gamma M,(\gamma+1)M)$, we have that the number of
deadlines from all packets on $e$ during $[t,t+T)$ is at most $T$ with
probability at least $1/2$.
\end{proof}

\paragraph{Remarks.}
To prove Lemma~\ref{l:whp} a condition weaker than~(\ref{eq:beta})
would be sufficient.  It would suffice to show that the number of
deadlines on any $e$ during any $[t,t+T)$ is at most
$(1+\ve/2)\sum_{e_0}\beta^\gamma_{e_0,e}T$.  Indeed, this would even
allow $T$ and $M$ to be a factor of $m$ smaller, as in
\cite{AndrewsAFKLL96}.  However, such a
weaker condition only allows derandomization in a centralized manner.

We emphasize that the condition~(\ref{eq:beta}) depends only on sets
of packets that are injected into {\em one particular} initial link.
Therefore we can choose the deadlines for a packet simply by
considering the other packets that are injected at the same initial
link.  Hence, we can carry out a derandomization independently at each
initial link and obtain a {\em distributed}, deterministic protocol.  This
is in contrast to the randomized protocol of
\cite{AndrewsAFKLL96} in which the success condition depends on
packets that are injected across all initial links in the network.

\paragraph{Derandomization.}
We use the method of conditional expectations to derandomize the
protocol for each $M$-interval.  (See e.g.~\cite{Raghavan88}.)
In summary,
\begin{theorem}
\label{t:derandomization}
Our derandomized protocol is distributed and guarantees a delay bound
of $2M=poly(m,w,1/\ve)$ for every packet.
\end{theorem}
\begin{proof}
Let
$S_{e_0,e}^{\gamma}=\{ p_0,p_1,\ldots,p_{\ell}\}$.  For $i\le \ell$,
let $g(\delta_0,\delta_1,\ldots,\delta_i)$ be equal to
{\footnotesize
\begin{eqnarray*}
\sum_{e,t}Pr\left[\sum_{p\in S_{e_0,e}^{\gamma}} X^{p,e}_{[t,t+T)} 
> (1+\ve/2)\beta^\gamma_{e_0,e}T |
\tau^{p_0}_0=\delta_0, \ldots, \tau^{p_i}_0=\delta_i\right],
\end{eqnarray*}
}where $t$ is summed over the range $[\gamma M, (\gamma+1)M-T)$.  By a
calculation similar to the Chernoff calculation
of~(\ref{eq:chernoff}), the value of $g(\cdot,\ldots,\cdot)$ is upper
bounded by the following function $h$,
\begin{eqnarray*}
h(\delta_0,\delta_1,\ldots,\delta_i) 
& = &\sum_{e,t}\frac{\prod_{p} \exp(\frac{\ve}{2}
E[X^{p,e}_{[t,t+T)} |
\tau^{p_0}_0=\delta_0,\ldots,\tau^{p_i}_0=\delta_i])}
{(1+\ve/2)^{(1+\ve/2)\beta^\gamma_{e_0,e}T}}.
\end{eqnarray*}

For fixed $\delta_0,\ldots,\delta_{i-1}$, the definition of
conditional expectation implies that there exists
an initial deadline $\delta_i$ for the packet $p_i$ such that
$h(\delta_0,\delta_1,\ldots,\delta_{i-1})\ge
h(\delta_0,\delta_1,\ldots,\delta_{i-1},\delta_i)$.  If we always
choose the initial deadline so that this inequality is satisfied
then,
\begin{eqnarray*}
g(\delta_0,\delta_1,\ldots,\delta_{\ell}) &\le&
h(\delta_0,\delta_1,\ldots,\delta_{\ell}))\\
&\le& h(\emptyset) \\
&\le& \exp(-\ve^2\beta^\gamma_{e_0,e}T/12),
\end{eqnarray*}
The third inequality follows from~(\ref{eq:prob}).  We have
chosen the parameters $M$ and $T$ so that 
$\exp(-\ve^2\beta^\gamma_{e_0,e}T/12)$ is less
than $1$.  In addition, since
$g(\delta_0,\delta_1,\ldots,\delta_{\ell})$ involves no randomness
every term of $g$ is either $0$ or $1$.  The above inequalities
imply that $g(\delta_0,\delta_1,\ldots,\delta_{\ell})$ is less than $1$ and so
condition~(\ref{eq:beta}) fails with probability zero.  Hence, with
probability one all deadlines are met and all packets reach their
destinations in time $2M$.

It remains to show that we can calculate
$h(\delta_0,\ldots,\delta_i)$.  If $j\le i$ then,
$$
E[X^{p_j,e}_{[t,t+T)} |
\tau^{p_0}_0=\delta_0,
\ldots,
\tau^{p_i}_0=\delta_i] 
$$
is equal to $0$ or $1$ depending on whether or not the initial
deadline $\delta_j$ causes packet $p_j$ to have a deadline for link
$e$ during $[t,t+T)$. If $j>i$ then,
$$
E[X^{p_j,e}_{[t,t+T)} |
\tau^{p_0}_0=\delta_0,
\ldots,
\tau^{p_i}_0=\delta_i] =
E[X^{p_j,e}_{[t,t+T)}],
$$
which is equal to the probability, over all possible choices of the
initial deadline, that packet $p_j$ has a deadline for link $e$ during
the interval $[t,t+T)$.  (Recall that the initial deadline has at most $M$
choices and all subsequent deadlines are chosen deterministically.)
This probability is solely dependent on whether or not the path for
packet $p_j$ passes through link $e$.  Hence, for fixed
$\delta_0,\ldots,\delta_{i-1}$ we can choose the value of $\delta_i$
that minimizes $h(\delta_0,\delta_1,\ldots,\delta_{i-1},\delta_i)$.
\end{proof}

\begin{figure}
\centerline{\input{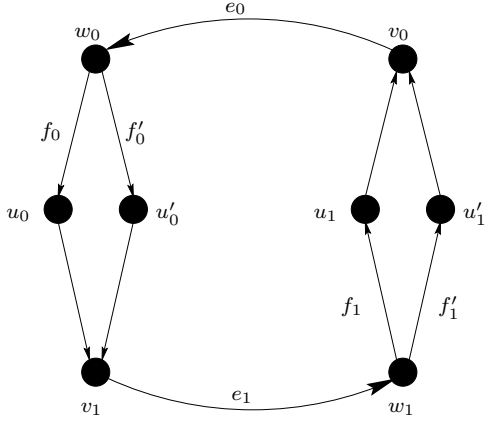}}
\caption{Network $G$ for which FIFO and NTG are unstable
even if we are allowed to choose routes.}
\label{f:ring}
\end{figure}
\section{Instability in Combined Routing and Scheduling}

In \cite{AndrewsAFKLL96} it was shown that if the packet routes are
given by the adversary then the FIFO and Nearest-to-Go (NTG)
scheduling protocols can be unstable even if the packet paths are
admissible.  (FIFO always gives priority to the packet that arrived at
the link earliest.  NTG always gives priority to the packet that has
the smallest number of hops remaining to its destination.)  However,
the examples given in \cite{AndrewsAFKLL96} do not lead to instability
if we are allowed to route packets on paths other than the ones chosen
by the adversary.

We therefore have a natural question.  If we are allowed to choose the
routes, can we guarantee that FIFO and NTG are stable? In this section
we show that the answer to this question is negative.  We present
examples in which regardless of how we choose the routes, the FIFO and
NTG scheduling protocols create instability.  


\begin{theorem}
There exists a network $G$ such that FIFO creates instability under
some $(w,r)$-admissible injections regardless of how packets are
routed.
\label{thm:unstable}
\end{theorem}
\begin{proof} 
Network $G$ is shown in Figure~\ref{f:ring}.  We break the packet
injections into phases.  We inductively assume that at the beginning
of phase $j$ a set $S$ of $s$ packets with destination $u_0$ is in the
queue of $e_0$.  We show that at the beginning of phase $j+1$ more
than $s$ packets with destination $u_1$ are in the queue of $e_1$.  By
symmetry this process repeats indefinitely and the number of packets
in the network grows without bound.  For the basis of the induction, we
inject a large burst of packets at source node $v_0$ with destination
node $u_0$, which is allowed by a large window $w$.  From now on all
the injections are at rate $r$ with burst size one.  In general the
sequence of injections in phase $j$ is as follows.
\begin{itemize}
\item[(1)] 
For the first $s$ steps, we inject a set $X$ of $rs$ packets at node
$v_0$ with destination $u_1$.  These packets are completely held up at
$e_0$ by the packets in $S$.  We also hold up packets in $S$ at $f_0$
by injecting $rs$ packets at $w_0$ with destination $u_0$.  These
newly injected packets get mixed with those of $S$ into the set $S'$.
At the end of the first $s$ steps, $rs$ packets from $S'$ are at
$f_0$. Note that packets in $X$ will be routed through either $f_0$ or
$f'_0$.
   
\item[(2)] 
For the next $rs$ steps, we inject a set $Y$ of $r^2s$ packets at node
$v_0$ with destination $u_1$. These packets are held up at $e_0$ by
the packets in $X$.  We also inject packets at $w_0$ with destination
$u'_0$ at rate $r$.  These packets delay the packets from $X$ that are
routed through $f'_0$.  Hence, at most $rs/(r+1)$ packets of $X$ cross
$f_0'$. (This only happens if packets in $X$ are routed through
$f'_0$, which is not necessarily the case.) Note that no packet from
$X$ crosses $f_0$ in these steps, since the packets in $S'$ have
priority. Hence, at the end of these $rs$ steps, a set $X'\subseteq X$
of at least $r^2 s/(r+1)$ packets are still at $w_0$.
   
\item[(3)] 
For the next $\ab{X'} + \ab{Y}$ steps the packets in $X'$ and $Y$ move
forward, and merge at $v_1$. Meanwhile, we inject packets at $v_1$
with destination $u_1$ at rate $r$. We end with at least $r(\ab{X'} +
\ab{Y})$ packets at $v_1$ with destination $u_1$. This number is at
least $r^3 s + r^3 s/(r+1)$.  
\end{itemize} 
This ends phase $j$.  For $r\ge 0.9$ we have $r^3+r^3/(r+1) > 1$.  It
is easy to verify that the injections during phase $j$ are admissible.
The inductive step is complete.
\end{proof}

Injections similar to the above can be used to prove
the instability of NTG on 
network $G$ at any rate $r>1/\sqrt{2}$.  The induction hypothesis of
phase $j$ now does not require the packets in $S$ to be initially in
the queue of $e_0$, but to cross $e_0$ in the first $s$ steps of the
phase.  Hence, subphase (3) is no longer required. Furthermore, after
subphase (2) both sets $Y$ and $X'$ contain at least $r^2 s$ packets,
since single-link injections have higher priority than the packets in
$X$.  It follows that the system is unstable since $2r^2s>s$.

\section{Stability of a Ring with Parallel Links}
\label{s:ring-parallel}

In this section we consider source routing on a ring with $c$ parallel
links. Consider a decomposition of the network into $c$ disjoint
single rings.  
We propose 
a deterministic on-line source-routing algorithm that routes
each packet along one of these rings and guarantees that the routing
is admissible. In \cite{AndrewsAFKLL96} it was
shown that the single ring is stable under any {\em greedy} scheduling
policy (i.e.\ one that always schedules a packet whenever packets are
waiting).  Hence, we conclude that the ring with $c$ parallel links is
stable under any greedy scheduling policy if our source-routing
algorithm is used.

Note that the 4-ring with 2 parallel links was shown to be unstable
under a greedy protocol such as FIFO when the packet paths are given
by the adversary~\cite{AndrewsAFKLL96}. This shows that freedom of
routing can make a difference in network stability since we have a
network that is unstable under FIFO if the adversary can dictate the
routes but is stable under FIFO if we can choose the routes intelligently.

\subsection{Definitions}
\label{a:ring}

Consider a ring with $n$ nodes and $c$ parallel directed links from
node $i$ to node $i+1 (\mbox{ mod } n)$.  The parallel links
connecting neighboring nodes are uniquely labeled $1,
\ldots, c$. We denote the cycle of $n$ links labeled $j$ as the $j$th
single ring. Note that, if $j \ne j'$, the $j$th and the $j'$th single
rings are link disjoint.  We assume that the injections are
$(w,r)$-admissible. For convenience we sometimes denote $1-r$ by
$\ve$. We propose a source-routing algorithm that finds weakly
$(W,R)$-admissible paths along these single rings, where,
\begin{eqnarray}
\label{eq-w}
W & = &\left\lceil\frac{3}{r\ve^2}\ln \frac{nc}{\beta}\right\rceil, \\
R & = & 1-\ve^2,
\end{eqnarray}
for some $\beta < 1$.

\subsection{Randomized Algorithm}

Let us first study the following randomized routing algorithm. Each
time a packet is injected, one of the $c$ single rings is randomly
chosen, uniform and independently, and the packet is routed along
it. 
Since the injections are $(w,r)$-admissible, in any $W$-interval at most
$crW$ packets are injected that must cross the parallel links from any
node $i$ to $i+1 (\mbox{ mod }n)$. Hence, the expected number of
packets routed along any link of the ring is at most $rW$. Using a
Chernoff bound we can upper bound the probability of more than
$(1+\ve)rW = RW$ packets being routed along any link in the
$W$-interval. Let $P=p_0,p_1,\ldots,p_{\ell}$ be the set of packets
injected in a $W$-interval. For each packet $p_j$, let $X_e^{p_j}$ be
the random variable denoting whether $p_j$ is routed along link
$e$. Let $X_e$ be the number of packets routed along link $e$ in the
$W$-interval. From a Chernoff bound we have that,
\begin{eqnarray*}
\label{eq-pr}
Pr[X_e > (1+\ve)rW] 
 & \le & \frac{\prod_{p_j \in P} E[(1+\ve)^{X_e^{p_j}}]}{(1+\ve)^{(1+\ve)rW}}\\
 & \le & [e^\ve / (1+\ve)^{(1+\ve)}]^{rW} \\
 & \le & e^{(\ve - (1+\ve) \ln (1+\ve))rW} \\
 & \le & (e^{-\ve^2/3})^{rW} \\
 & \le & \frac{\beta}{nc}.
\end{eqnarray*}
The last two inequalities follow from the fact that $\ve<1$ and the
definition of $W$ in~(\ref{eq-w}), respectively. We can now bound the
probability of {\em any} link having more than $(1-\ve^2)W$ packets
routed along it. We use $E$ to denote the set of links in the ring.
\begin{eqnarray*}
\label{eq-prmax}
Pr[\max_{e \in E} X_e > (1+\ve)rW] 
          & \le & \sum_{e \in E} Pr[X_e > (1+\ve)rW] \\
          & \le & |E| \frac{\beta}{nc} \\
          & = & \beta
\end{eqnarray*}
Hence, since $\beta < 1$, there is a positive probability of routing
all the packets in such a way that no link has congestion more than
$RW$. By choosing a very small $\beta$ (e.g., $O(1/n)$) we
could show that this randomized algorithm guarantees that the routing
is weakly $(W,R)$-admissible with high
probability. This can be used to show the stability of any greedy
scheduling protocol in a probabilistic sense (i.e., there is a value
$C$ such that the probability of having more than $kC$ packets in the
system at any given time is exponentially small in $k$).

However, in the rest of the section we only need $\beta < 1$. We will
derandomize the proposed algorithm, and all we need for this process
to work is to have a feasible routing with the required
properties. This is guaranteed for any $\beta < 1$.

\subsection{Off-line Routing}

We will now derandomize the above algorithm so that all the packets
are deterministically routed and no link has congestion more than
$(1-\ve^2)W$. To do this, we use the method of conditional
probabilities, as we did in Section~\ref{s:polydelay}. Unfortunately, 
to apply this method directly we need to
know from the beginning the set $P$ of packets to be routed. 
We achieve this as follows.
We divide time into intervals of $W$ steps, and
hold all the packets injected in one $W$-interval until its last
step. Then, all these packets are routed in that last step, when all
of them are known.

Let $P=p_0,p_1,\ldots,p_{\ell}$ be the set of packets injected in a
$W$-interval. Let $\gamma_{p_j}$ denote the single ring chosen to
route packet $p_j$.  For $i\le \ell$ let, $$
g(\delta_0,\delta_1,\ldots,\delta_i) = Pr[\max_{e \in E} X_e >
(1+\ve)rW | \gamma_{p_0}=\delta_0,\ldots,\gamma_{p_i}=\delta_i].$$
Since $g(\cdot,\ldots,\cdot)$ is difficult to calculate directly, we
define another function $h(\cdot,\ldots,\cdot)$ by,
$$
h(\delta_0,\delta_1,\ldots,\delta_i) = 
\sum_{e \in E} \frac{\prod_{p_j \in P} E[(1+\ve)^{X_e^{p_j}}| 
\gamma_{p_0}=\delta_0,\ldots,\gamma_{p_i}=\delta_i]}{(1+\ve)^{(1+\ve)rW}},
$$ which can be easily computed. For this, it is enough so observe
that, when computing $h(\delta_0,\delta_1,\ldots,\delta_i)$, for each
packet $p_j$,
\begin{itemize}
\item
if $j \le i$, then 
\begin{itemize}
\item
if $e$ is in the $\delta_j$th single ring and it is in the
path from the source to the destination of $p_j$, then $E[(1+\ve)^{X_e^{p_j}}| 
\gamma_{p_0}=\delta_0,\ldots,\gamma_{p_i}=\delta_i] = 1+\ve$.
\item
Otherwise, $E[(1+\ve)^{X_e^{p_j}}| 
\gamma_{p_0}=\delta_0,\ldots,\gamma_{p_i}=\delta_i] = (1+\ve)^0=1$.
\end{itemize}
\item
if $j > i$, then 
\begin{itemize}
\item
if $e$ could be in the
path from the source to the destination of $p_j$, then 
$E[(1+\ve)^{X_e^{p_j}}|\gamma_{p_0}=\delta_0,\ldots,\gamma_{p_i}=\delta_i]=(1+\ve)^{1/c}$.
\item
Otherwise, $E[(1+\ve)^{X_e^{p_j}}| 
\gamma_{p_0}=\delta_0,\ldots,\gamma_{p_i}=\delta_i] = (1+\ve)^0=1$.
\end{itemize}
\end{itemize}

We have that,
$g(\delta_0,\delta_1,\ldots,\delta_i) \le
h(\delta_0,\delta_1,\ldots,\delta_i)$. Also, for fixed
$\delta_0,\ldots,\delta_{i-1}$, the definition of conditional
expectation implies that the single ring $\delta_i$ can be chosen such
that $h(\delta_0,\delta_1,\ldots,\delta_{i-1})\ge
h(\delta_0,\delta_1,\ldots,\delta_{i-1},\delta_i)$.  If we always
choose the single rings so that this inequality is satisfied then, $$
g(\delta_0,\delta_1,\ldots,\delta_{\ell}) \le
h(\delta_0,\delta_1,\ldots,\delta_{\ell})) \le h(\emptyset) \le
\beta.$$ In this expression, the left-hand-side involves no randomness
and so it is either $0$ or $1$.  However, since $\beta < 1$, it has to
be less than $1$ and so there must be a probability zero of failure.
Hence, no link has congestion more than $(1-\ve^2)W$, and the routing
is weakly $(W,R)$-admissible.

\subsection{On-line Routing}

Now we want to route packets as soon as they are injected. This does
not allow us to directly use the above derandomization process, since
we will not necessarily know the set $P$ by the time we need to route the
first packets. This is needed to compute the different values of the
function $h(\cdot,\ldots,\cdot)$.  However, we will deal with this
problem by making pessimistic assumptions about the packets that have
not been injected yet.

First consider two packets, $p_k$ and $p_l$, such that their paths do
not overlap, and the destination node of $p_k$ is the source node
of $p_l$. Replace these packets by one single packet whose source
node is that of $p_k$ and its destination node is that of $p_l$.
Observe that, for fixed $\delta_0,\ldots,\delta_i$, if $k > i$ and $l
> i$, the value of $h(\delta_0,\ldots,\delta_i)$ does not change by
the replacement (see above). This can be generalized to the
replacement of any number of packets.

Then, this allows us to use the following trick. Initially we assume a
set $P^{(0)}$ of packets that consists of $crW$ {\em ghost packets}
going from node $i$ to node $i+1 (\mbox{ mod }n)$, for each
$i$. The value $h(\emptyset)$ is computed for this set
$P^{(0)}$.

Now, assume that $i-1$ packets have been already injected and routed. (That
is, the values $\delta_0,\delta_1,\ldots,\delta_{i-1}$ are fixed and
$h(\delta_0,\delta_1,\ldots,\delta_{i-1})$ is computed.) When the
$i$th packet $p_i$ is injected, we remove one ghost packet from the
set $P^{(i-1)}$ for each hop that $p_i$ crosses.  These ghost packets
are replaced by the packet $p_i$ to obtain a new set $P^{(i)}$.
The existence of the appropriate ghost packets is guaranteed
by the initial ghost packets we put in $P^{(0)}$ and the fact that the
injections are  $(w,r)$-admissible. As we saw previously, this
does not change the value of
$h(\delta_0,\delta_1,\ldots,\delta_{i-1})$.  Then, route the packet
$p_i$ (choose and fix $\delta_i$) so that
$h(\delta_0,\delta_1,\ldots,\delta_{i-1})\ge
h(\delta_0,\delta_1,\ldots,\delta_{i-1},\delta_i)$.

By repeating this process, at the end of the $W$-interval we have that 
$$
g(\delta_0,\delta_1,\ldots,\delta_{\ell}) \le
h(\delta_0,\delta_1,\ldots,\delta_{\ell})) \le h(\emptyset) \le
\beta,$$ where $\ell$ is the number of packets injected during the
$W$-interval. We now remove all the remaining ghost packets. This process
eliminates any remaining randomness in
$g(\delta_0,\delta_1,\ldots,\delta_{\ell})$, and can never increase
its value, since it only removes packets. Then, since
$g(\delta_0,\delta_1,\ldots,\delta_{\ell})=0$ involves no randomness
and $\beta<1$, $g(\delta_0,\delta_1,\ldots,\delta_{\ell})=0$ and no link
has congestion more than $(1-\ve^2)W$. Hence, the routing is weakly
$(W,R)$-admissible.

\section{Conclusions}

In this paper we have presented source routing algorithms for
packet-switched networks and we have described the first distributed,
deterministic scheduling protocol with a polynomial delay bound.
There is much still to be explored in the study of combined routing
and scheduling.  For example, different packets are often associated
with different delay requirements.  Some of them may be
delay-sensitive whereas others may be delay-tolerant.  The problem of
scheduling these packets on given routes in order to meet these delay
requirements has been studied before.  The ability to choose the
routes would add an additional dimension to the problem and may even
make scheduling easier.

\subsection*{Acknowledgment}
The authors wish to thank Adam Meyerson for helpful discussions.


\begin{thebibliography}{10}

\bibitem{AielloKOR98}
{\sc W.~Aiello, E.~Kushilevitz, R.~Ostrovsky, and A.~Rosen}, {\em Adaptive
  packet routing for bursty adversarial traffic}, in Proceedings of the 30th
  Annual {ACM} Symposium on Theory of Computing, Dallas, TX, May 1998, pp.~359
  -- 368.

\bibitem{AndrewsAFKLL96}
{\sc M.~Andrews, B.~Awerbuch, A.~Fern\'andez, J.~Kleinberg, T.~Leighton, and
  Z.~Liu}, {\em Universal stability results and performance bounds for greedy
  contention-resolution protocols}, Journal of the ACM, 48 (2001), pp.~39--69.

\bibitem{AwerbuchAP93}
{\sc B.~Awerbuch, Y.~Azar, and S.~Plotkin}, {\em Throughput competitive on-line
  routing}, in Proceedings of the 34th Annual Symposium on Foundations of
  Computer Science, 1993, pp.~32--40.

\bibitem{AwerbuchAPW94}
{\sc B.~Awerbuch, Y.~Azar, S.~Plotkin, and O.~Waarts}, {\em Competitive routing
  of virtual circuits with unknown duration}, in Proceedings of the 5th Annual
  {ACM-SIAM} Symposium on Discrete Algorithms, 1994, pp.~321--330.

\bibitem{AwerbuchL94}
{\sc B.~Awerbuch and T.~Leighton}, {\em Improved approximation algorithms for
  the multicommodity flow problem and local competitive routing in dynamic
  networks}, in Proceedings of the 26th Annual {ACM} Symposium on Theory of
  Computing, 1994, pp.~487--496.

\bibitem{BertsimasG99}
{\sc D.~Bertsimas and D.~Gamarnik}, {\em Asymptotically optimal algorithm for
  job shop scheduling and packet routing}, Journal of Algorithms, 33 (1999),
  pp.~296--318.

\bibitem{BorodinKRSW96}
{\sc A.~Borodin, J.~Kleinberg, P.~Raghavan, M.~Sudan, and D.~P. Williamson},
  {\em Adversarial queueing theory}, Journal of the ACM, 48 (2001), pp.~13--38.

\bibitem{Gamarnik98}
{\sc D.~Gamarnik}, {\em Stability of adversarial queues via fluid models}, in
  Proceedings of the 39th Annual Symposium on Foundations of Computer Science,
  Palo Alto, CA, November 1998, pp.~60--70.

\bibitem{Gamarnik99}
\leavevmode\vrule height 2pt depth -1.6pt width 23pt, {\em Stability of
  adaptive and non-adaptive packet routing problems in adversarial queueing
  networks}, in Proceedings of the 31th Annual {ACM} Symposium on Theory of
  Computing, Atlanta, GA, May 1999, pp.~206--214.

\bibitem{GargK98}
{\sc N.~Garg and J.~{K\"onemann}}, {\em Faster and simpler algorithms for
  multicommodity flow and other fractional packing problems}, in Proceedings of
  the 39th Annual Symposium on Foundations of Computer Science, Palo Alto, CA,
  November 1998, pp.~300--309.

\bibitem{Keshav97}
{\sc S.~Keshav}, {\em An engineering approach to computer networking}, Addison
  Wesley, Reading, MA, 1997.

\bibitem{LeightonMR88}
{\sc F.~T. Leighton, B.~M. Maggs, and S.~B. Rao}, {\em Packet routing and
  job-shop scheduling in {O}(congestion + dilation) steps}, Combinatorica, 14
  (1994), pp.~167 -- 186.

\bibitem{pst:pack95}
{\sc S.~Plotkin, D.~Shmoys, and E.~Tardos}, {\em Fast approximation algorithms
  for fractional packing and covering problems}, Math of Oper. Research,
  (1994), pp.~257--301.

\bibitem{Raghavan88}
{\sc P.~Raghavan}, {\em Probabilistic construction of deterministic algorithms:
  approximating packing integer programs}, Journal of Computer and System
  Sciences, 37 (1988), pp.~130 -- 143.

\bibitem{RFC3031}
{\sc E.~Rosen, A.~Viswanathan, and R.~Callon}, {\em Multiprotocol label
  switching architecture}.
\newblock {RFC 3031}, 2001.
\newblock http://www.ietf.org/rfc/rfc3031.txt.

\bibitem{SrinivasanT97}
{\sc A.~Srinivasan and C.~Teo}, {\em A constant-factor approximation algorithm
  for packet routing, and balancing local vs. global criteria}, in Proceedings
  of the 29th Annual {ACM} Symposium on Theory of Computing, El Paso, TX, May
  1997, pp.~636 -- 643.

\bibitem{Braden}
{\sc D.~Tennenhouse, J.~Smith, W.~Sincoskie, D.~Wetherall, and G.~Minden}, {\em
  A survey of active network research}, IEEE Communications Magazine,  (1997),
  pp.~80--86.

\bibitem{y:round95}
{\sc N.~Young}, {\em Randomized rounding without solving the linear program},
  ACM-SIAM Symposium on Discrete Algorithms,  (1995), pp.~170--78.

\end{thebibliography}

\end{document}